\documentclass[
  reprint,
  amsmath,
  amssymb,
  aps,
  prl,
  superscriptaddress
]{revtex4-2}
\usepackage{graphicx} 
\usepackage{hyperref} 
\usepackage{physics}
\usepackage{pgf}
\usepackage{siunitx}
\begin{document}

\title{Measurement-Induced Quantum Neural Network}

\author{Paul Argyle}
\affiliation{Department of Physics, University of Maryland, College Park, Maryland 20742, USA}
\affiliation{Joint Quantum Institute, University of Maryland and National Institute of Standards and Technology, College Park, Maryland 20742, USA}

\author{Djamil Lakhdar-Hamina}
\affiliation{Department of Physics, University of Maryland, College Park, Maryland 20742, USA}
\affiliation{Joint Quantum Institute, University of Maryland, College Park, Maryland 20742, USA}

\author{Sarah H. Miller}
\affiliation{Applied Research Laboratory for Intelligence and Security, University of Maryland, College Park, Maryland 20742, USA}

\author{Victor Galitski}
\affiliation{Department of Physics, University of Maryland, College Park, Maryland 20742, USA}
\affiliation{Joint Quantum Institute, University of Maryland and National Institute of Standards and Technology, College Park, Maryland 20742, USA}

\date{\today}

\begin{abstract}
We introduce a measurement-induced quantum neural network (MINN), an adaptive monitored-circuit architecture in which mid-circuit measurement outcomes determine the entangling gates in subsequent layers. In contrast to standard monitored circuits where sites and gates are sampled randomly, the gates are parametrized and variational, producing correlated history-dependent dynamics and injecting nonlinearity through measurement back-action. A generic MINN is not expected to be efficiently classically simulable. To demonstrate feasibility, we study a matchgate MINN that admits exact fermionic simulation and can be trained with gradient estimators. We apply the architecture to continuous optimization, image classification, and ground-state search in the Sherrington-Kirkpatrick spin glass, finding effective training and performance over a broad range of monitoring rates. 
\end{abstract}

\maketitle

Quantum machine learning has attracted attention as a near-term application of noisy intermediate-scale quantum (NISQ) devices, due to its potential to operate on noisy hardware with limited qubit count \cite{Biamonte2017,Cerezo2021,Havlicek2019}. In parallel, the extension of quantum computing ideas to deep learning has produced an extensive literature on quantum neural networks (QNNs)\cite{Kak1995,Schuld2014,farhi2018classificationquantumneuralnetworks,Beer2020}. The most common approach defines a QNN as a variational quantum circuit: qubits serve as neurons, unitary transformations act as hidden layers, and the output is obtained through measurement. Learning is performed by minimizing the expectation value of a task-dependent cost function. However, because such circuits often contain no intrinsic hidden-layer nonlinearity, it is unclear whether they should be described as neural networks. Moreover, it remains unclear how the nonlinear expressive capacity central to classical neural networks is realized in these putative quantum counterparts.

In this work, we introduce a novel quantum neural network architecture, which we call a Measurement-Induced Neural Network (MINN). Inspired by brick-wall circuit geometries commonly studied in measurement-induced phase transitions (MIPT) \cite{MIPT2018,MIPT2019,MIPT2019Nahum}, our architecture is instead adaptive: measurement results in a layer determine rotation angles in the next layer. The MINN introduces nonlinearity through measurement back-action and adaptive feed-forward between layers and therefore this architecture provides a natural mechanism for incorporating hidden-layer nonlinearity. 
A generic MINN is not  classically simulable due to the exponential growth of the underlying Hilbert space.
In this work, we did not have access to quantum hardware capable of implementing a fully generic MINN, and thus restricted gates to a simulable subset of $SU(4)$. We use several representative benchmarking problems: (i) minimization of testbed objective functions, (ii) image classification, and (iii) ground-state search for the Sherrington--Kirkpatrick (SK) spin glass.

\begin{figure*}[t]
\centering
\includegraphics[height=0.45\textheight]{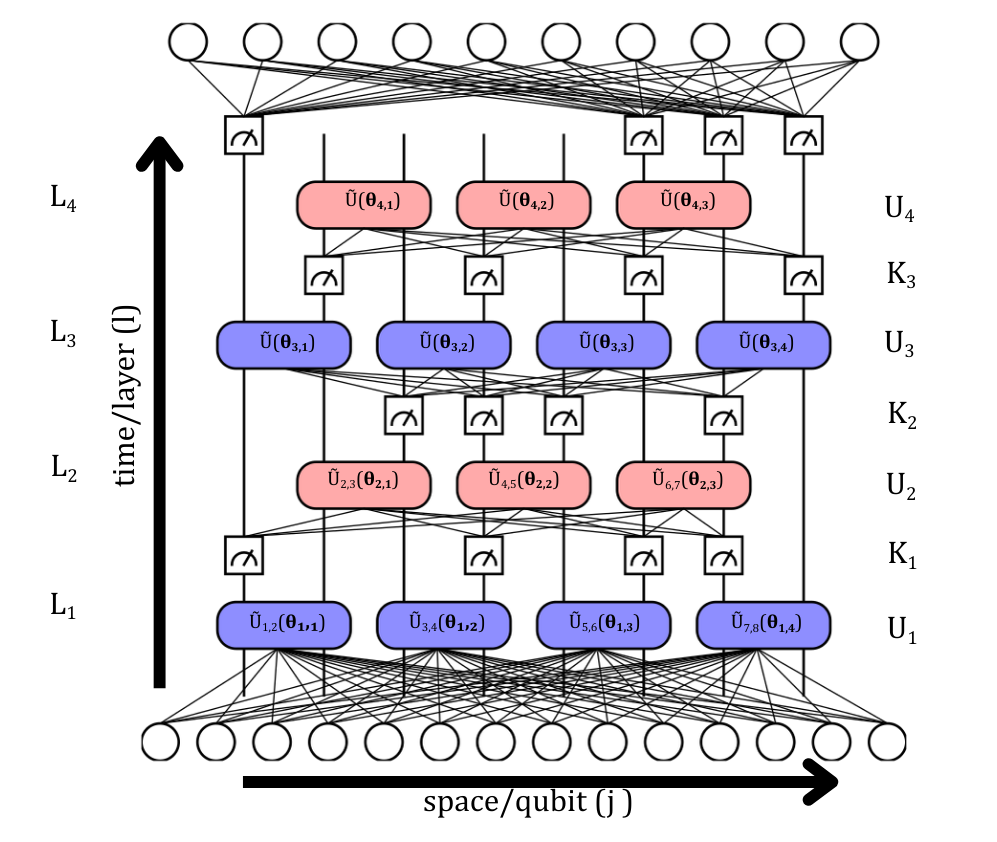}
\caption{
Brick-wall circuit architecture commonly studied in MIPT. Sites are labeled by spacetime coordinates $(l,j)$; here the circuit contains four layers and eight qubits. The alternating even–odd pattern of two-qubit gates spreads entanglement throughout the system. Gate subscripts indicate both the qubits acted upon and the layer index of the first qubit in each pair, illustrated explicitly in the first two layers. Rotation angles $\boldsymbol{\theta}$ subscripts indicate the time index and the first qubit index. Projective measurements punctuate the spacetime circuit, producing alternating unitary and measurement layers. The MINN extends this architecture to a dynamic and adaptive circuit in which measurement outcomes feed forward to subsequent unitary layers, while circuit parameters , the rotation angles, are optimized to minimize the expectation value of a loss function.
}
\label{fig:nn_fig}
\end{figure*}

The circuit is defined on a $(1+1)$-D spacetime lattice with
time layers $l=1,\ldots,L$ and qubit indices $j=1,\ldots,N$, labeling
sites $(l,j)$. Each layer acts on nearest-neighbor pairs in a
brick-wall geometry depicted in Fig. \ref{fig:nn_fig}.


We define the hidden-layer dynamics as a layered,
parameterized quantum instrument with classical feedforward which,
at the ensemble level, induces a completely positive trace-preserving
(CPTP) map
$
\rho(\mathcal{W})=\mathcal{E}_{\mathcal{W}}(\rho_0),
$
acting on an input state of \(N\) qubits encoded as a computational-basis
bitstring, $\rho_0 = |x\rangle\langle x|.$
The channel $\mathcal{E}_{\mathcal W}$ is composed of $L$ alternating
unitary and measurement layers. Operationally, the channel is implemented by sampling quantum
trajectories. The ordered set
of measurement outcomes, called a record, 
\[
\boldsymbol{\tau}
=
\bigl(\boldsymbol{\mu}^{\,1},\boldsymbol{\mu}^{\,2},\ldots,\boldsymbol{\mu}^{\,L}\bigr),
\qquad
\boldsymbol{\mu}^{\,l}
=
\{\mu^l_j\}_{j\in \pi_l},
\]
labels a quantum trajectory where $\pi_l \subset \{1,\dots,N\}
$.

For a trajectory labeled by the measurement record
$\boldsymbol{\tau}$, the corresponding completely positive map is
\[  
\mathcal{E}_{\mathcal{W},\boldsymbol{\tau}}
=
\mathcal{L}_{L,\boldsymbol{\mu}^{\,L}}
\circ
\mathcal{L}_{L-1,\boldsymbol{\mu}^{\,L-1}}
\circ
\cdots
\circ
\mathcal{L}_{1,\boldsymbol{\mu}^{\,1}} ,
\]
where $\mathcal{L}_{l,\boldsymbol{\mu}^{\,l}}$ is a channel corresponding to one layer of gates and mid-circuit measurements. For unitary layers the map acts as
\[
\mathcal{U}^{(\pi_l)}_{l,\boldsymbol{\tau}_{<l}}(\rho)
=
U^{(\pi_l)}_{l}(\mathcal W,\boldsymbol{\tau}_{<l})
\,\rho\,
U^{(\pi_l)\dagger}_{l}(\mathcal W,\boldsymbol{\tau}_{<l}),
\]
where $\boldsymbol{\tau}_{<l}$ denotes the measurement record prior to
layer $l$. The unitary layers are factorized into a brick-wall geometry
of even and odd bonds,
\[
\begin{aligned}
U_l^{(\mathrm{odd})}(\mathcal W,\boldsymbol{\tau}_{<l})
&=
\bigotimes_{k=1}^{\lfloor N/2 \rfloor}
\widetilde U_{2k-1,\,2k}
\!\bigl(\theta_{l,2k-1}(\mathcal W,\boldsymbol{\tau}_{<l})\bigr),\\[6pt]
U_l^{(\mathrm{even})}(\mathcal W,\boldsymbol{\tau}_{<l})
&=
\bigotimes_{k=1}^{\lfloor (N-1)/2 \rfloor}
\widetilde U_{2k,\,2k+1}
\!\bigl(\theta_{l,2k}(\mathcal W,\boldsymbol{\tau}_{<l})\bigr).
\end{aligned}
\]
 To enable efficient classical simulation, we restrict the dynamics (the $\widetilde{U}$) to the matchgate subgroup of $SU(4)$, corresponding to the quadratic fermionic algebra obtained via the Jordan–Wigner transformation. Each two–qubit gate is of the form $\widetilde{U}=e^{-iH}$ with Hamiltonian
\begin{equation}
 \begin{aligned}
H = \frac{1}{2}(&\theta_{XX}\, X \otimes X
+ \theta_{XY}\, X \otimes Y
+ \theta_{YX}\, Y \otimes X \\
&+ \theta_{YY}\, Y \otimes Y
+ \theta_{ZI}\, Z \otimes I
+ \theta_{IZ}\, I \otimes Z )
\end{aligned}
\label{eq:matchgate_pauli_basis}
\end{equation}
Gates generated by Hamiltonians of this form are known as matchgates \cite{Valiant2001,DiVincenzo,Bravyi2005FLO}. A brief introduction to matchgates is provided in Appendix~\ref{app:matchgates}, and the full matchgate simulation code is available at ~\cite{brickwallnet2026}.

The parameters $\boldsymbol{\theta}^{\,l}$ of the gates at layer $l$ are generated by a classical function whose inputs are the measurement outcomes from the previous layer. Denoting the measurement record by $ \boldsymbol{\mu}^{\,l}
=
(\mu^{\,l}_1,\mu^{\,l}_2,\ldots,\mu^{\,l}_N),$
the rotation parameters are computed as
\begin{equation}
\boldsymbol{\theta}^{\,l}
= \frac{\pi}{2}\Bigl(\boldsymbol{1} - \phi_a\!\left(\mathbf{W}^{\,l-1}\boldsymbol{\mu}^{\,l-1} + \mathbf{b}^{\,l-1}\right)\Bigr).
\label{eq:theta_map}
\end{equation}
In Eq.~\eqref{eq:theta_map}, the outcomes for sites that are not measured are defined to be zero; for $ j \notin \pi_l,\mu^l_j\equiv0 $. This is analogous to dropout in classical neural networks \cite{Srivastava2014}.

Here $\mathbf{W}^{\,l}$ denotes the trainable weight matrix at layer $l$, and $\mathbf{b}^{\,l}$ the bias vector. The weights are initialized using Kaiming initialization \cite{he2015delving} and the biases used in mapping measurements to unitary layers are initialized to zero. We use different initializations for the input and output layer biases for different problems. The weights and biases constitute the variational parameters of the model, which determine the rotation angles in $SU(4)$ through the nonlinear function $\phi_a$.

The function $\phi_a(x)$ is applied element-wise and defined here by
\begin{equation}
\phi_a(x)=
\mathrm{htanh}\!\left(\frac{x}{a}\right)
=
\begin{cases}
\dfrac{x}{a}, & |x| \le a, \\[6pt]
\operatorname{sgn}(x), & |x| > a .
\end{cases}
\end{equation}
The parameter $a$ controls the width of the activation function and therefore the degree of nonlinearity in the parameter map. When gates reduce to single-qubit $X$ or $Y$ rotations with $p=1$ and
$a\to0$, the model becomes a classical partially binarized neural
network. In this case $a$ interpolates between classical ($a=0$) and quantum
($a>0$) regimes \cite{barney2025naturalquantizationneuralnetworks}.

Measurement layers correspond to projective measurements
performed with probability $p$ on sites $j\in\pi_l$. More generally,
$p$ may be site-dependent.

The measurement channel satisfies the normalization condition
\[ (1-p)I + p\sum_{\mu=-1,1} P^{(\mu)}_j = I,
\]
where
\[
P^{(\mu)}_j
=
\frac{1}{2}\left(I + \mu Z_j\right),
\qquad
\mu\in\{-1,1\},
\]
are computational-basis projectors acting on qubit $j$.

Operationally, the measurement layer is implemented by sampling
measurement outcomes $\boldsymbol{\mu}^{\,l}$,
which define the trajectory-dependent Kraus operator
\[
K_l(\boldsymbol{\mu}^{\,l})
=
\prod_{j\in\pi_l} P^{(\mu^l_j)}_j .
\]

Conditioned on a measurement record
$\boldsymbol{\tau}$,
the circuit produces the unnormalized trajectory state
\[
\widetilde{\rho}_{\boldsymbol{\tau}}
=
K_L(\boldsymbol{\mu}^{\,L})U_L\cdots
K_1(\boldsymbol{\mu}^{\,1})U_1
\,\rho_0\,
U_1^\dagger K_1^\dagger(\boldsymbol{\mu}^{\,1})
\cdots
U_L^\dagger K_L^\dagger(\boldsymbol{\mu}^{\,L}).
\]

The probability of observing trajectory $\boldsymbol{\tau}$ is
\[
p_{\mathcal{W}}(\boldsymbol{\tau})
=
\mathrm{Tr}\!\left(\widetilde{\rho}_{\boldsymbol{\tau}}\right),
\]
and the normalized conditional state is
\[
\rho_{\boldsymbol{\tau}}
=
\frac{\widetilde{\rho}_{\boldsymbol{\tau}}}{p_{\mathcal{W}}(\boldsymbol{\tau})}.
\]

Only the measurement record $\boldsymbol{\tau}$ enters the parameter map
in Eq.~\eqref{eq:theta_map}. Concretely, at each spacetime site $(l,j)$, we sample a Bernoulli random variable that determines whether a measurement occurs. With probability $p$, a projective measurement is performed on the qubit at that location.
One can fix the measurement sites across trajectories or reset sites per trajectory. 

Given the  record of measurement outcomes $\boldsymbol{\mu}^{L}$ at the final layer $L$, the MINN maps these outcomes to a classical output layer. 
The precise form of this output layer depends on the task.
When output variables are continuous, the final activations are linear
\begin{equation}
\mathbf{x}_{\mathrm{out}}
= \mathbf{W}^{L} \boldsymbol{\mu}^{L} + \mathbf{b}^{L}.
\label{eq:linear_input}
\end{equation}
In the discrete case, when the output variables are binary $x_{\mathrm{out},i} = \pm 1$, the network output specifies the probability that each degree of freedom - for our SK example interpreted as a spin - occupies a given state. The probability that the network predicts $x_{\mathrm{out},i} = +1$ is
\[P(x_{\mathrm{out},i}=+1)
=
\sigma\!\left(
\sum_j W^{L}_{ij}\, \mu^{L}_j + b^{L}_i
\right).
\]
In order to perform inference, the feedforward, trajectory sampling, must be executed multiple times to obtain an estimate of the output. For our classification task, we apply the sigmoid function to the linear output Eq.~\eqref{eq:linear_input} to obtain a logit for each class. We average logits over inference shots. The final prediction is the label corresponding to the largest average logit.

Classical neural networks are typically trained by minimizing a task-dependent cost function defined over the activations of the final layer. Analogously, for each problem we define a cost function on measurement outcomes $C(\boldsymbol{\tau})$. The function $C(\boldsymbol{\tau})$ can take the form of any standard cost function with activations of the last hidden layer replaced with the corresponding measurement results $\boldsymbol{\mu}^{L}$. The MINN is trained by minimizing the average cost
\begin{equation}
\mathbb{E}_{\boldsymbol{\tau} \sim p_\mathcal{W}} \Bigl[ C(\boldsymbol{\tau}) \Bigr] = \sum_{\boldsymbol{\tau}} p_{\mathcal{W}} (\boldsymbol{\tau}) C(\boldsymbol{\tau}),
\label{eq:expC}
\end{equation}

where $p_{\mathcal{W}}(\boldsymbol{\tau})$ is the probability the network outputs measurement register $\boldsymbol{\tau}$ for a given value of the weights and biases. We minimize the average cost with stochastic gradient descent using the REINFORCE gradient estimator \cite{Williams1992REINFORCE}. This is obtained by differentiating Eq.~\eqref{eq:expC} with respect to trainable parameters $\mathcal{W}$ and applying the log-derivative trick,
\begin{equation}
\nabla_{\mathcal{W}}  \mathbb{E}_{\boldsymbol{\tau} \sim p_\mathcal{W}} \Bigl[ C(\boldsymbol{\tau}) \Bigr]
=
\mathbb{E}_{\tau \sim p_\mathcal{W}}\!\left[S_{\mathcal{W}}(\boldsymbol{\tau})\,C(\boldsymbol{\tau})\right],
\label{eq:reinforce}
\end{equation}
\[
S_{\mathcal{W}}(\boldsymbol{\tau})
=
\nabla_{\mathcal{W}} \log p_{\mathcal{W}}(\boldsymbol{\tau}).
\]
The quantity $S_{\mathcal{W}}(\boldsymbol{\tau})$ is called the score. We discuss how to efficiently calculate the score for the matchgate MINN in Appendix~\ref{app:score_calculation}. Because REINFORCE estimators can suffer from high variance,
we introduce a per-parameter baseline $B$ to reduce variance
\cite{greensmith2004variance},
\begin{equation}
\nabla_{\mathcal{W}}  \mathbb{E}_{\boldsymbol{\tau} \sim p_\mathcal{W}} \Bigl[ C(\boldsymbol{\tau}) \Bigr]
=
\mathbb{E}_{\boldsymbol{\tau} \sim p_\mathcal{W}}\!\left[
S_{\mathcal{W}}(\boldsymbol{\tau})\,
\bigl(C(\boldsymbol{\tau})-B\bigr)
\right].
\label{eq:grad_estimator}
\end{equation}

The optimal baseline for variance reduction is
\begin{equation}
B_{\mathrm{opt}}
=
\frac{ \mathbb{E}_{\boldsymbol{\tau} \sim p_\mathcal{W}} \Bigl[  C(\boldsymbol{\tau}) S_{\mathcal{W}}^{2}(\boldsymbol{\tau}) \Bigr] 
}{
\mathbb{E}_{\boldsymbol{\tau} \sim p_\mathcal{W}} \Bigl[ S_{\mathcal{W}}^{2}(\boldsymbol{\tau}) \Bigr]
}.
\end{equation}

We now show results from applying the MINN to three tasks: (i)  minimization of test-bed objective functions, (ii) image classification, and (iii) ground-state search for the SK spin glass. The set-up and hyper-parameters for all our simulations are given in Table~\ref{tab:hyperparam}.

We apply the neural network architectures to the task of minimizing continuous cost functions. The benchmark functions exhibit rugged landscapes which make them non-trivial test-beds for optimization. Although this is not a practical application domain for neural networks, these tests provide a controlled setting in which to evaluate performance across non-trivial loss landscapes. 

\begin{figure}[t]
\centering
\includegraphics[width=0.95\linewidth]{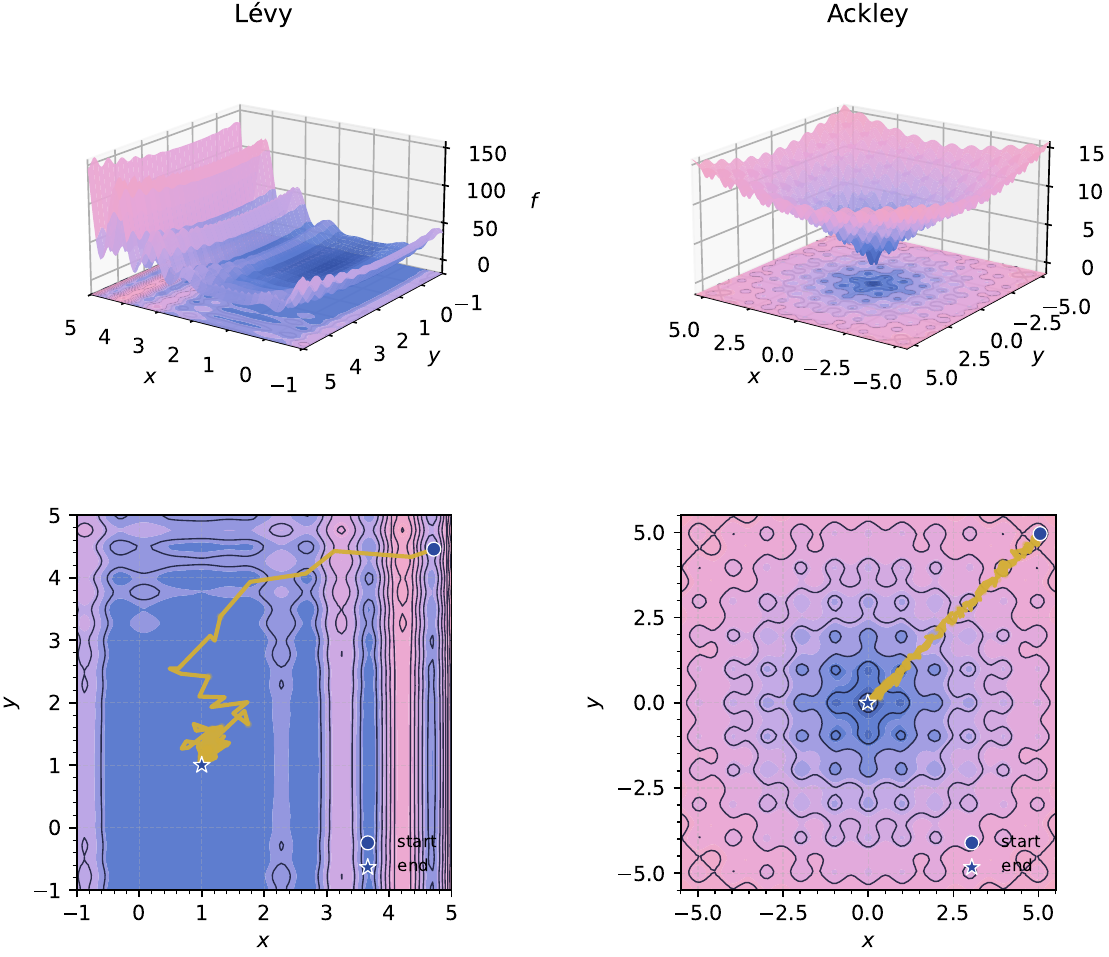}
\caption{Optimization landscapes for the Lévy (top left) and Ackley (top right) benchmark functions. Bottom panels display the optimization trajectories projected onto contour maps, illustrating convergence toward the global minima; trajectories are highlighted in gold. The measurement probability is $p=0.5$ with measurement sites reset between evaluations.}
\label{fig:landscapes_w_trajectories}
\end{figure}

\begin{figure}[t]
\centering
\includegraphics[width=0.95\linewidth]{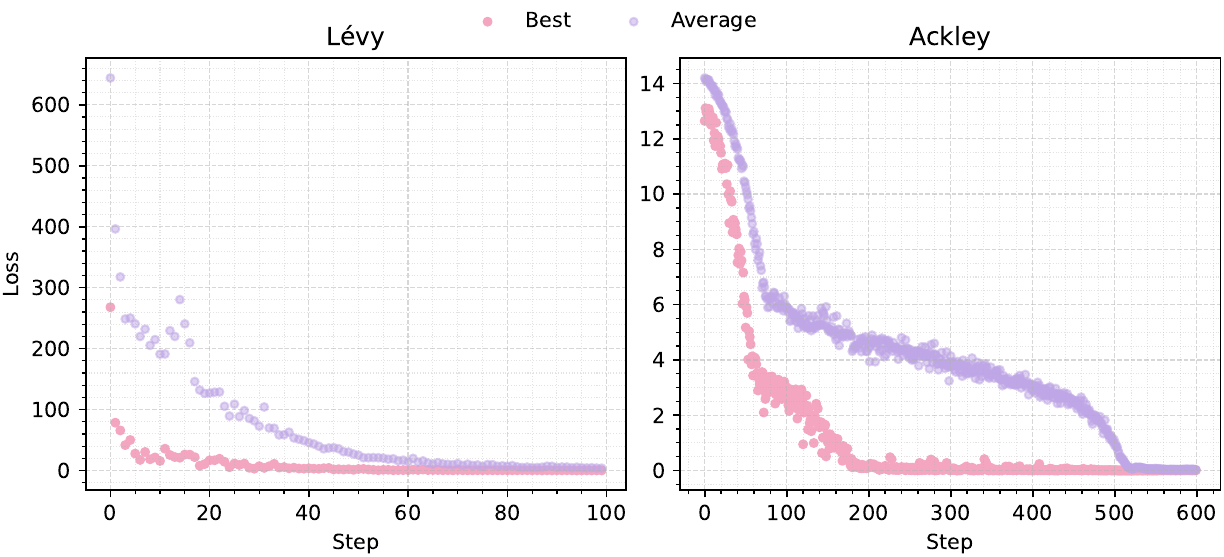}
\caption{For a 10-d case of each loss function: loss as a function of optimization step, showing both the best value obtained within each sampling batch and the sample-averaged (coarse-grained) loss. Left is Lévy and right Ackley. The measurement probability is $p=0.5$ with measurement sites reset each time the circuit is evaluated during training or inference.}
\label{fig:landscapes}
\end{figure}

We consider two standard optimization test functions: the Lévy function Eq. \ref{eq:levy_loss}, which contains numerous local minima and a narrow basin surrounding the global minimum; and the Ackley function Eq. \ref{eq:ackley_loss}, characterized by a nearly flat outer region combined with rapidly oscillating local minima \cite{Surjanovic2013VLSE}. These functions map an $n$-dimensional vector $\mathbf{x}$ to a scalar $C$ and are therefore defined in arbitrary dimensions. Fig.~\ref{fig:landscapes_w_trajectories} shows the objective functions together with the action of coarse-grained (multi-sample-per-step) stochastic gradient descent in their 2-D forms. Fig.~\ref{fig:landscapes} shows the coarse-grained mean loss trajectories, as well as the best per sample trajectory, for each test function in 10-D. In both cases, the networks converge toward the global minimum.
\begin{equation}
\begin{aligned}
C_1(\boldsymbol{x})&= \sin^2(\pi x_1)
+ \sum_{i=1}^{n-1} (x_i-1)^2
\Bigl[1 + 10\sin^2(\pi x_i + 1)\Bigr] \\
&\quad + (x_n-1)^2
\Bigl[1 + \sin^2(2\pi x_n)\Bigr].
\end{aligned}
\label{eq:levy_loss}
\end{equation}

\begin{equation}
\begin{aligned}
C_2(\boldsymbol{x}) &= -20 \exp\!\left(
-0.2 \sqrt{\frac{1}{n}\sum_{i=1}^n x_i^2}
\right) \\
&\quad - \exp\!\left(
\frac{1}{n}\sum_{i=1}^n \cos(2\pi x_i)
\right) + 20 + \exp(1).
\end{aligned}
\label{eq:ackley_loss}
\end{equation}

\begin{figure}[t]
\centering
\resizebox{\linewidth}{!}{\input{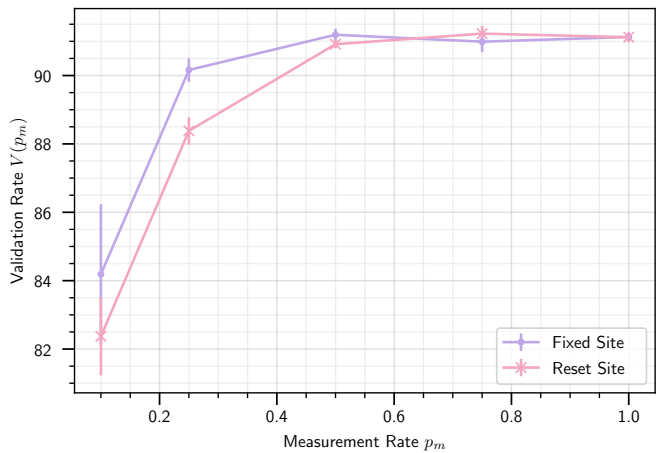}}
\caption{Inference results on MNIST show that validation performance is unaffected by measurement sites being fixed or reset. Near-unitary evolution or moderate nonlinearities lead to improved classification accuracy. Measurement plays a role analogous to dropout, indicating redundancy in the learned representation.}
\label{fig:mnist}
\end{figure}

Next, we tackle a typical machine-learning benchmark: image classification on the MNIST dataset \cite{mnist}. The MNIST dataset consists of 60,000 $28 \times 28$  pixel black and white images used for training and 10,000 images for testing \cite{LeCun1998}. However, to test regularization capacities, we train on 5,000 images. The network is trained using batched stochastic gradient descent \cite{Robbins1951}. The gradient is calculated by averaging the gradient estimator Eq.~\eqref{eq:grad_estimator} $N_{\text{samples}}=200$ with the data input for each sample randomly drawn from the batch. The model is trained using cross-entropy loss. We train and run inference using both fixed and non-fixed measurement locations. The validation rate on the training set for each case at different values of the measurement probability $p$ are shown in Fig.~\ref{fig:mnist}. 
We see that the network performs effectively for a range of $p<1$. The validation rate does not substantially decrease until around $p\approx0.1$. We do not expect critical $p$ because we are in a finite-size regime. The continuous crossover, however, does seem to show that measurement exposes redundancy of information. 

Finally, we apply the matchgate neural network to a discrete optimization problem of physical relevance. The goal is to find the ground state of the Sherrington-Kirkpatrick Spin-Glass Hamiltonian: 
\[H = - \sum_{i>j} J_{ij} s_i s_j,\quad J_{ij}\!\sim\!\mathcal{N}(0,1/N).
\]
where $s_i = \pm1$ and the couplings $J_{ij}$ are  independent identically distributed Gaussian entries of zero mean and variance $1/N$.

The Sherrington–Kirkpatrick (SK) model is an all-to-all Ising spin glass that exhibits frustration and quenched disorder in its couplings \cite{mezard1987spin}. Determining its' ground state is equivalent to solving a weighted max-cut problem on a complete graph and therefore in the worst-case belongs to the NP-hard complexity class \cite{montanari2021sk}. The computational difficulty arises from a rugged free-energy landscape populated by numerous metastable states and characterized, in the mean-field solution, by an ultrametric (hierarchical) organization of states \cite{mezard1987spin,rammal1986ultrametricity}. Consequently, methods for finding ground-state energies rely on heuristic algorithms in the general case.

\begin{table}[t]
\label{tab:hyperparam}
\begin{ruledtabular}
\begin{tabular}{l c c c c}
Hyperparameter 
& \shortstack{Ackley$_{10\text{D}}$}
& \shortstack{Levy$_{2\text{D},10\text{D}}$,\\Ackley$_{2\text{D}}$} 
& MNIST 
& Max-Cut \\
\hline
Qubits & 32 & 32 & 64 & 100 \\
Meas.\ Prob.\  & 0.5 & 0.5 & \textemdash & 0.5 \\
Hidden Layers  & 2 & 2 & 2 & 4 \\
Step Size  & 0.01 & 0.005 & 0.01 & 0.01 \\
Momentum  & 0.9 & 0.5 & 0.9 & 0.95 \\
Grad.\ Samples  & 100 & 100 & 200 & 100 \\
Infer.\ Samples  & 100 & 100 & 200 & 100 \\
Width $a$  & 1.0 & 1.0 & 3.0 & 1.0 \\
Epochs  & \textemdash & \textemdash & 50 & \textemdash \\
Batch Size & \textemdash & \textemdash & 70 & \textemdash \\
Training Size & \textemdash & \textemdash & 5000 & \textemdash\\
Test Size & \textemdash & \textemdash & 10000 & \textemdash\\
Init. In Bias ($b^0_i$) & $\sim \mathcal{U}(-a,a)$ & $\sim \mathcal{U}(-a,a)$ & $ 0$ & $\sim \mathcal{U}(-a,a)$ \\
Init. Out Bias ($b^L_i$) &  5 & 5 & 0 & 0 \\
\end{tabular}
\end{ruledtabular}
\caption{
Hyperparameters used for each benchmark task. The Lévy 2-D and 10-D and Ackley 2-D functions share the same hyperparameters within each dimensional setting, as indicated. Step size and momentum apply to the stochastic gradient descent with momentum algorithm \cite{Goodfellow2016}. ``Grad.\ Samples'' and ``Infer.\ Samples'' denote the number of trajectory samples used to estimate gradients and perform inference, respectively. The parameter $a$ controls the width of the activation function. Biases are initialized either to zero or drawn from a uniform distribution $\mathcal{U}(-a,a)$, as specified. Dashes indicate that the corresponding parameter is not applicable to the task.
}
\end{table}

\begin{figure}[t]
\centering
\includegraphics[width=\linewidth]{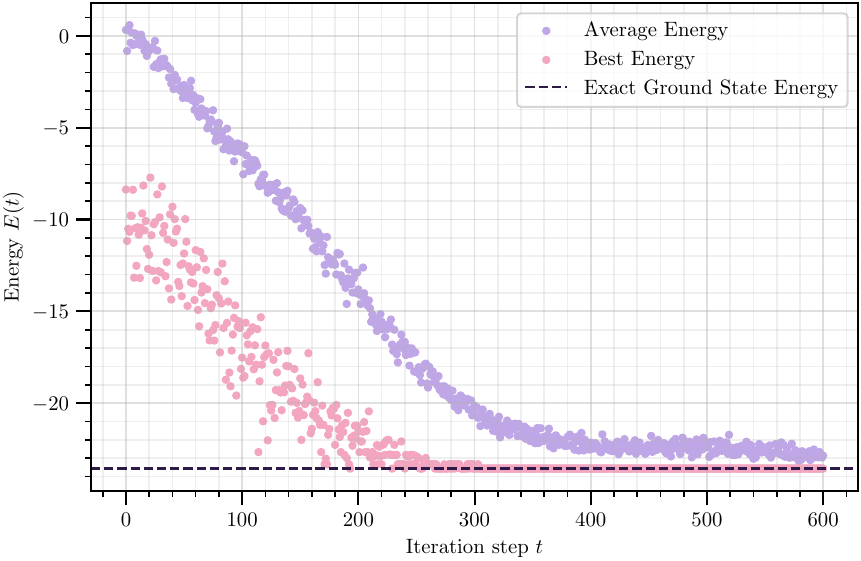}
\caption{Energy as a function of optimization step. The MINN is effectively optimized, with the ground-state energy estimate converging smoothly during training. The measurement probability is $p=0.5$ with measurement sites reset between evaluations.}
\label{fig:energy_vs_step}
\end{figure}

We minimize the Hamiltonian for a single realization of the couplings $J$; the specific instance used in this study is provided in Ref.\cite{brickwallnet2026}. The network may be interpreted as a parameterized pdf over spin configurations: each output node defines the probability that a spin is in the up or down state, and the joint output defines a probability over configurations. Training seeks parameter values that increase the probability weight assigned to configurations with energies arbitrarily close to the ground state. The lowest sampled energy and the sample-averaged energy predicted during training are shown in Fig.~\ref{fig:energy_vs_step}. The system consists of 36 spins. We benchmark our results against an exact ground-state solution obtained through the use of mixed-integer linear programming \cite{gass1985linear}. 


In this work, we proposed a novel quantum neural network architecture. The key distinction between MINN and typical adaptive variational networks is punctuation by mid-circuit measurements, i.e., we elevate the hidden-layer dynamics from unitary time evolution to an open quantum system. We find that MINN solves instances of different classes of problems. Simulations were restricted to the class of matchgates. These networks were shallow, with at most four hidden layers. We believe shallow networks will not exhibit dynamics characteristic of deeper generic MINN i.e. only in a thermodynamic-like regime do monitored \textit{deep} brick-wall circuits exhibit measurement induced phase transitions. 


The natural sequel to this work is to implement a deep MINN with the full $SU(4)$ gate set on quantum hardware and solve problems where it has been shown that classical methods and existing quantum architectures struggle. Although conventional training methods such as parameter shift or simultaneous perturbation stochastic approximation can be applied to a general MINN, they require a large number of shots due to the high variance of these gradient estimators \cite{schuld2019gradient}. More efficient training methods should be developed. Typical parameterized quantum circuits suffer from the barren plateau problem, where gradients vanish exponentially with system size \cite{McClean2018}. Future work should investigate whether the barren plateau problem can be avoided in the MINN. At sufficiently high measurement rate $p$, gradients no longer decay exponentially, but this may come at the cost of expressivity \cite{Wiersema2023,Gyawali2024}. We are working to implement deep MINN on actual hardware, improve training, and analyze the training landscape of the MINN. 

{\em Acknowedgements --} The authors are grateful to Dr.~David Hayes and the Quantinuum team for providing access to their quantum computing hardware for preliminary experimental testing of key concepts related to the measurement-induced neural network. This work was supported by the US Army Research Office under Grant Number W911NF-23- 1-024 and Simons Foundation (P.J.A. and V.G.).
\bibliographystyle{apsrev4-2}
%


\newpage
\setcounter{secnumdepth}{1} 
\appendix
\section{Matchgate Simulation}
\label{app:matchgates}
In this section we give an overview of the method for classically simulating the 1D matchgate brickwall circuit. For more information on the classical simulation of matchgate and gaussian fermionic systems see \cite{Valiant2001,DiVincenzo,Bravyi2005FLO}. Matchgates can be viewed as separate $SU(2)$ rotations on the even parity sub-space ($\text{span}\{ \ket{00},\ket{11} \}$) and odd parity subspace ($ \text{span} \{ \ket{01} , \ket{10} \}$). A general matchgate $F(A,B)$ can be expressed as 
\begin{equation}
F(A,B) = \begin{pmatrix} A_{11} & 0 & 0 & A_{12} \\ 0 & B_{11} & B_{12} & 0 \\ 0 & B_{21} & B_{22} & 0 \\ A_{21} & 0 & 0 & A_{22} \end{pmatrix}.
\end{equation}
where
\[A = \begin{pmatrix} A_{11} & A_{12} \\ A_{21} & A_{22} \end{pmatrix}, B = \begin{pmatrix} B_{11} & B_{12} \\ B_{21} & B_{22} \end{pmatrix} \]
with $A,B \in SU(2)$.
Matchgates form a six-parameter family isomorphic of $SU(2) \oplus SU(2)$. A general matchgate admits a Pauli-basis expansion Eq.~\eqref{eq:matchgate_pauli_basis}. Circuits initialized in a tensor product state and evolving via nearest neighbor matchgates interspersed with mid-circuit projective Pauli-$Z$ measurements can be simulated classically in polynomial time. The $N$-qubit density matrix can be mapped to an $N$-mode fermionic Gaussian state. Both nearest-neighbor matchgates and projective Pauli-$Z$ measurements are Gaussian channels, i.e., they map Gaussian states to Gaussian states. An $N$-mode Gaussian fermionic state is completely specified by its $2N \times 2N$ covariance matrix. The matchgate simulation reduces to updating the covariance matrix of the corresponding Gaussian fermionic system.

We represent operators on an $N$-qubit system with $2N$ Majorana operators in the Jordan-Wigner representation, 
\begin{equation}
c_{2j-1} = Z^{\otimes(j-1)}X_j I^{\otimes(N-j)}, c_{2j} = Z^{\otimes(j-1)}Y_j I^{\otimes(N-j)},
\end{equation}
for $j=1,...,N.$ The Majorana operators satisfy the Clifford algebra
\begin{equation}
\{c_\alpha, c_\beta\} = c_\alpha c_\beta + c_\beta c_\alpha = 2 \delta_{\alpha \beta} I \qquad \alpha,\beta =1,\dots,2N.
\end{equation}
Expressing a nearest-neighbor matchgate Hamiltonian Eq.~\eqref{eq:matchgate_pauli_basis} in terms of the Majorana operators
\begin{equation}
H_{j,j+1} = i \sum_{\alpha, \beta =2j-1}^{2j+2} h_{\alpha \beta} c_\alpha c_\beta
\end{equation}
where $h$ is the anti-symmetric matrix
\begin{equation}
h= \frac{1}{4}\begin{pmatrix}
0 & -\theta_{ZI} & \theta_{YX} & \theta_{YY} \\
\theta_{ZI} & 0 & -\theta_{XX} & -\theta_{XY} \\
-\theta_{YX} & \theta_{XX} & 0 & -\theta_{IZ} \\
-\theta_{YY} & \theta_{XY} & \theta_{IZ} & 0
\end{pmatrix}
\label{eq:hmatrix}
\end{equation}
 The Hamiltonian of nearest-neighbor matchgates is quadratic in the Jordan-Wigner basis. Consider the evolution of a Majorana operator under a quadratic Hamiltonian,
 \begin{equation}
 c_\alpha(t) = e^{i Ht} c_\alpha(0) e^{-i H t}, \qquad H = i\sum_{ \beta, \beta^\prime} h_{ \beta \beta^\prime} c_\beta c_{\beta^\prime}.
 \end{equation}
The clifford algebra and Heisenberg eqns of motion yield
\begin{equation}
\frac{d c_\alpha (t)}{dt} = \sum_\beta 4 h_{\alpha \beta} c_\beta (0).
\end{equation}
Setting $t=1$ and suppressing labels of time dependence,
\begin{equation}
e^{i H}c_\alpha e^{-iH} =  \sum_\beta R_{\alpha \beta} c_\beta,
\end{equation}
where $R \in SO(2N)$ is given by $R=e^{4h}$. Consider a system in a Gaussian state with respect to the Majorana operators. Such a state can be represented as 
\begin{equation}
\rho = \frac{1}{Z} \exp \Bigl( -\frac{i}{4} \sum_{\alpha \beta} K_{\alpha \beta} c_\alpha c_\beta \Bigr),
\end{equation}
where $K_{\alpha \beta}$ is a real anti-symmetric matrix (the generator matrix) and $Z$ is a normalization factor. Under quadratic evolution,
\begin{equation}
U \rho U^\dagger = \frac{1}{Z} \exp \Bigl( -\frac{i}{4} \sum_{\alpha \beta}K^\prime_{\alpha \beta} c_\alpha c_\beta \Bigr) 
\end{equation}
where $K^\prime = R K R^T$. Thus, nearest-neighbor matchgates in 1D constitute a Gaussian channel. The Gaussian state is fully characterized by its covariance matrix $\Gamma$ defined by
\begin{equation}
\Gamma_{\alpha \beta} = \frac{i}{2} \Tr\Bigl( \rho [c_\alpha,c_\beta] \Bigr) =  \begin{cases}
    i \Tr \Bigl( \rho c_\alpha c_\beta \Bigr),   & \text{if } \alpha \neq \beta \\
    0,   & \text{if } \alpha = \beta
\end{cases}.
\end{equation}
Under the action of a quadratic Hamiltonian, the covariance matrix evolves as
\begin{equation}
\Gamma \rightarrow R \Gamma R^T.
\end{equation}
Projective Pauli-$Z$ measurements also define a gaussian channel \cite{Bravyi2005FLO}. We do not review the proof here, but we discuss the corresponding update rule for the covariance matrix. Let $P_j^{(\mu)}$ denote the projector onto the $Z_j$-eigenstate with outcome $\mu= \pm 1$. In the Majorana basis,
\begin{equation}
P_j^{(\mu)} = \frac{1}{2}\Bigl(1-i \mu c_{2j-1} c_{2j} \Bigr). 
\end{equation}
The post-measurement state ${\rho^\prime}_j^{(\mu)}$ in terms of the pre-measurement state $\rho$ is
\begin{equation}
{\rho^\prime}^{(\mu)}_j = \frac{P_j^{(\mu)} \rho P_j^{(\mu)}}{p_j^{(\mu)}},
\end{equation}
where $p_j^{(\mu)} = \text{Tr}\Bigl(P_j^{(\mu)} \rho \Bigr)$ is the measurement probability. The updated two-point correlator,  $(G^{\prime(\mu)}_j)_{\alpha \beta } = \Tr \Bigl( {\rho^\prime}_j^{(\mu)} c_\alpha c_\beta \Bigr)$ is
\[(G^{\prime(\mu)}_j)_{\alpha \beta}  = \frac{1}{2 p_j^{(\mu)}} \langle c_\alpha c_\beta \rangle - \frac{i \mu}{ 2p_j^{(\mu)}} \langle c_\alpha c_\beta c_{2j-1} c_{2j} \rangle, \]
where the expectation values $\langle \dots\rangle$ are taken with respect to the pre-measurement state. The system is initialized in a Gaussian state and evolves under Gaussian channels, so the pre-measurement state is Gaussian. Thus the four-point correlator can be expressed in terms of two-point correlators using Wick's theorem,
\begin{equation}
(G^{\prime(\mu)}_j)_{\alpha \beta}=G_{\alpha \beta} -\frac{i \mu}{2 p_j^{(\mu)}} \Bigl( G_{\alpha 2j} G_{\beta 2j-1} - G_{\alpha 2j-1} G_{\beta 2j}  \Bigr).
\end{equation}
The corresponding update for the covariance matrix is
\begin{equation}
 \Gamma^{\prime(\mu)} = \Gamma +  \Gamma A^{(\mu)}_j \Gamma  + A^{(\mu)}_j ,
 \label{eq:meas_update}
 \end{equation}
 \[(A^{(\mu)}_j)_{\alpha \beta} \equiv \frac{\mu}{2p_j^{(\mu)}}  \Bigl(\delta_{\alpha 2j} \delta_{\beta 2j-1}- \delta_{\beta 2j} \delta_{\alpha 2j-1}\Bigr).\]
The measurement probability in terms of the pre-measurement covariance matrix,
\begin{equation}
 p_j^{(\mu)} = \frac{1}{2}\Bigl( 1 - \mu \Gamma_{2j-1,2j} \Bigr).
 \label{eq:meas_prob}
\end{equation}
 Our MINN simulations are initialized in the product state $\ket{0}^{\otimes N}$ with $\Gamma = I_N \otimes (-i \sigma_y)$. For each layer $l=1,\dots,L$, the covariance matrix is updated via $R^l\in SO(2N)$.\\
For odd $l$,
\begin{equation}
R^l = R_{2q}(\theta^l_1) \oplus R_{2q}(\theta^l_2) \oplus...\oplus R_{2q}(\theta^l_{N/2}).
\end{equation}
For even $l$,
\[R^l =  I_{2} \oplus R_{2q}(\theta^l_1) \oplus ... \oplus R_{2q}(\theta^l_{N/2-1}) \oplus I_2. \]
Here $R_{2q}=e^{4h}$ denotes the two-qubit matrix obtained by exponentiating Eq.~\eqref{eq:hmatrix}. After each unitary layer, stochastic measurement updates Eq.~\eqref{eq:meas_update} are iteratively applied with probabilities given by Eq.~\eqref{eq:meas_prob}. In the MINN measurement results determine rotation angles in the subsequent layer.

\section{Matchgate Score Calculation}
\label{app:score_calculation}
We describe our procedure for calculating the score entering the gradient estimate  Eq.~\eqref{eq:grad_estimator}. The method involves forward and backward propagation steps, analogous to reverse-mode differentiation in classical neural networks. The feed-forward step consists of simulating the circuit as discussed in Appendix~\ref{app:matchgates} and recording all measurement results, pre-measurement covariance matrices, and angles. For a sampled trajectory $\tau$, the log-lieklihood $\log p(\tau)$ decomposes as a sum of contributions, each depending only on the covariance matrix immediately prior to a measurement. This permits efficient backpropagation of the log-likelihood with respect to the covariance matrix. We derive these backpropagation maps for both measurement operations and unitary layers, together with the expression relating the score at each layer to the gradient of the log-likelihood with respect to the corresponding covariance matrix. 

The circuit has $L$ brickwall layers, with $M^l$ measurements in layer $l$. We denote by $\Gamma^l_0$ the covariance matrix immediately before the $l$th unitary layer, and by $\Gamma^{l}_i (i=1,\dots, M^l)$ the covariance matrix immediately prior to the $i$th measurement in that layer. The corresponding measurement probability, measurement site, and outcome are denoted $p^l_i$, $j^l_i$, and $\mu^l_i$, respectively. 
The probability of a quantum trajectory $\tau$ factorizes as
\begin{equation}
p(\tau) = \prod_{l=1}^L \prod_{i=1}^{M^l} p^l_i .
\end{equation}
Taking the logarithm of $p(\tau)$ and expressing the conditional probabilities in terms of covariance matrices,
\[\log p(\tau) = \frac{1}{2} \sum_{l=1}^L \sum_{i=1}^{M^l}  \Bigl( 1 - \mu^l_i ({\Gamma^l_{i})_{2j^l_i-1 2j^l_i}} \Bigr).  \]
Differentiating $\log p(\tau)$ with respect to the covariance matrix just prior to the last measurement,
\begin{equation}
\frac{\partial \log p(\tau)}{\partial \Gamma^l_{M^L}}= \frac{\partial}{\partial \Gamma^L_{M^L}} \log \Bigl(1+\mu^L_{M^L} (\Gamma^L_{M^L})_{2 j^L_{M^L} 2j^L_{M^L}-1 }\Bigr) = A^L_{M^L},
\end{equation}
\[(A^l_i)_{\alpha \beta} \equiv \frac{\mu^l_i}{2 p^l_i}(\delta_{2j^l_i \alpha } \delta_{2j^l_i-1 \beta} -\delta_{2j^l_i \beta} \delta_{2j^l_i -1 \alpha}).\]
Given $\partial \log p(\tau)/\partial \Gamma^l_{i+1}$ we can backpropagate through the measurement map to obtain the gradient with respect to $\Gamma^l_i$,
\begin{equation}
\frac{\partial \log p(\tau)}{\partial (\Gamma^l_i)_{\alpha \beta}} = \sum_{\rho \sigma} \frac{\partial\log p(\tau)}{\partial (\Gamma^l_{i+1})_{\rho \sigma} } \frac{\partial (\Gamma^l_{i+1})_{\rho \sigma}}{\partial (\Gamma^l_i)_{\alpha \beta}} 
\end{equation}
\[+ \frac{\partial}{\partial (\Gamma^l_i)_{\alpha \beta}} \log \Bigl(1+\mu^l_{i} (\Gamma^l_i)_{2 j^l_{i} 2j^l_{i}-1 }\Bigr).\]
Differentiating Eq.~\eqref{eq:meas_update} yields the measurement backpropagation relation
\begin{equation}
\frac{\partial \log p(\tau)}{\partial \Gamma^l_i} = \frac{\partial \log p(\tau)}{\partial \Gamma^l_{i+1}} +\frac{\mu^l_i}{2p^l_i} \Bigl(Q^l_i \frac{\partial \log p(\tau)}{\partial \Gamma^l_{i+1}} + \frac{\partial \log p(\tau)}{\partial \Gamma^l_{i+1}} ({Q^l_i})^T \Bigr)
\end{equation}
\[-\frac{1}{2 p^l_i} \Bigl(\Gamma^l_i \frac{\partial \log p(\tau)}{\partial \Gamma^l_{i+1}} \Gamma^l_i+ \frac{\partial \log p(\tau)}{\partial \Gamma^l_{i+1}}   \Bigr)_{2j^l_i,2j^l_i-1} A^l_i + A^l_i ,\]
where $(Q^l_{i})_{\alpha \beta} \equiv \delta_{2j^l_i \alpha} (\Gamma^l_i)_{2j^l_i-1 \beta} - \delta_{2j^l_i-1 \alpha} (\Gamma^l_i)_{2j^l_i \beta}$. This relation can be iterated to backpropagate through all measurements in a layer. Subsequently, we calculate the score for that layer. The gradient with respect to some angle $\theta^l$ in the layer is
\begin{equation}
\frac{\partial \log p(\tau)}{\partial \theta^l} = \frac{1}{2} \text{Tr} \Bigl(\frac{\partial \log p(\tau)}{\partial \Gamma^l_1} \frac{\partial \Gamma^l_1}{\partial \theta^l} \Bigr). 
\end{equation}
Expressing this in terms of the $SO(2N)$ matrix $R^l$ discussed in Appendix~\ref{app:matchgates},
\begin{equation}
\frac{\partial \log p(\tau)}{\partial \theta^l} = \frac{1}{2} \text{Tr} \Bigl(\frac{\partial \log p(\tau)}{\partial \Gamma^l_1} \Bigl[\frac{\partial R^l}{\partial \theta^l} (R^{l})^T, \Gamma^l_1 \Bigr]  \Bigr).
\label{eq:theta_grad}
\end{equation}
The matrix $\partial R^l / \partial\theta^l$ is only non-zero in the $4 \times 4$ block corresponding to $\partial R_{2q}/\partial \theta^l$ for the appropriate gate. We calculate $\partial R_{2q}/\partial \theta^l$ numerically using the Fr\'{e}chet derivative \cite{Rudin1991}. We obtain $\partial \log p(\tau)/\partial W^{l-1}$ and $\partial \log p(\tau)/ \partial b^{l-1}$ from Eq.~\eqref{eq:theta_grad} and Eq.~\eqref{eq:theta_map}.

After calculating the score for a layer we backpropagate through the unitary layer using
\begin{equation}
\frac{\partial \log p(\tau)}{\partial \Gamma^l_0} = ({R^l})^T \frac{\partial \log p(\tau)}{\partial \Gamma^l_1}R^l.
\end{equation}
This process continues backwards through all layers.

\label{matchgates}
\end{document}